\documentclass[lineno]{JFM-FLM_Au}
\usepackage{natbib}

\lefttitle{I. Addison-Smith, I.A. Maia, A.V.G. Cavalieri and B. Herrmann}
\righttitle{Journal of Fluid Mechanics}

\title{Mean-flow-based reduced-order models of turbulent channel flow}

\author{Ian Addison-Smith\aff{1}, Igor A. Maia\aff{2}, André V. G. Cavalieri\aff{2}  \and Benjamin Herrmann\aff{3,4}}

\affiliation{\aff{1}Department of Mechanical Engineering, Universidad de Chile, Beauchef 851, Santiago, Chile
\aff{2}Divisão de Engenharia Aeronáutica, Instituto Tecnológico de Aeronáutica, São José dos Campos
12228-900, Brazil
\aff{3}Department of Mechanical and Metallurgical Engineering, Pontificia Universidad Católica de Chile, Av. Vicuña Mackenna 4860, Santiago, Chile
\aff{4}Department of Hydraulic and Environmental Engineering, Pontificia Universidad Católica de Chile, Av. Vicuña Mackenna 4860, Santiago, Chile
}

\corresau{Ian Addison-Smith, \email{ian.addisonsmith@ug.uchile.cl}}

\begin{document}
\maketitle

\begin{abstract}
Reduced-order models (ROMs) for turbulent flows based on Galerkin projection can achieve reasonable accuracy using equation-based modal bases derived from the linearized Navier-Stokes equations through the controllability and observability Gramians. The use of the modal bases obtained from linearized equations around a mean state has been seen to enhance the first- and second-order statistics in the ROM, but the use of the mean state was not necessarily extended to the equations of motion, as it implies the treatment of the divergence of the Reynolds stresses in the Galerkin projection. In this work, we present a mean-flow-based framework for ROMs in which the projection of the Reynolds stresses is solved through a modified modal basis and the knowledge of the mean flow. This framework achieves turbulence statistics comparable to those of a reference direct numerical simulation (DNS) in a minimal channel at $Re_{\tau} \approx 185$. Short-time forecasting with this framework is assessed, where balanced  truncation modal bases outperform controllability modes in ROMs, yielding a reconstruction of the velocity field comparable to the Galerkin projection of proper orthogonal decomposition (POD) modes. This framework can extend analysis based on linearisations around the mean turbulent flow, which became widespread in recent years, to include explicitly non-linear interactions between modes, enabling accurate models at higher Reynolds number.
\end{abstract}

\begin{keywords}
% Authors should not enter keywords on the manuscript, as these must be chosen by the author during the online submission process and will then be added during the typesetting process (see \href{https://www.cambridge.org/core/journals/journal-of-fluid-mechanics/information/list-of-keywords}{Keyword PDF} for the full list).  Other classifications will be added at the same time.
\end{keywords}

%{\bf MSC Codes }  {\it(Optional)} Please enter your MSC Codes here

\section{Introduction}
\label{sec:headings}

Turbulent flows are characterized by strongly nonlinear and chaotic dynamics, where the dimensionality of the dynamics is often high for typical configurations of interest. Reduced-order models (ROMs) seek to model the fluid flow in a lower dimensional space \citep{rowley2017arfm}, achieving significant computational savings and simplified flow dynamics. ROMs have successfully addressed multiple challenges of fluid flow, such as  modeling the dynamics of coherent structures (e.g. \citealt{aubry1988jfm, waleffe1997pof, moehlis2004njp, cavalieri2021prf, mccormack2024jfm}), flow control applications (e.g. \citealt{ilak2008pof, barbagallo2009jfm, bagheri2009amr, dergham2011pof, maia2025jfm}), and forecasting the flow field (e.g. \citealt{noack2003jfm,rowley2005ijbc, nakamura2021pof, linot2023jfm}).

A common approach in ROMs is to use Galerkin projection, where the Navier-Stokes equations are projected onto a suitable modal basis. Proper orthogonal decomposition (POD) is a standard data-driven modal decomposition \citep{berkooz1993arfm} used in reduced-order modeling. It determines a subspace that is optimal in the sense of maximizing the captured kinetic energy, obtained by solving an eigenvalue problem using snapshot data from simulations or experiments. Given a low-rank expansion of the velocity field using these modes, the governing Navier–Stokes equations are projected onto the modal basis. This ROM is commonly referred to as POD-Galerkin and has been applied to different flow configurations over time (e.g. \citealt{aubry1988jfm, noack2003jfm, noack2005jfm, smith2005jfm, khoo2022jfm, sato2025jfm}). However, it has been argued that neglecting higher-order POD modes causes models to suffer from numerical instability due to the truncation of relevant non-energetic interactions present in the turbulent energy cascade. Latest works show that modal bases from the linearized Navier-Stokes equations in Galerkin ROMs (e.g. \citealt{cavalieri2022prf}) can overcome the instability without needing a closure for the truncated terms as other works (\citealt{protas2015jfm, hijazi2020jcp, ahmed2021pof, khoo2022jfm}). 

Unlike POD modes, which require flow data to construct the basis, modal decompositions derived from the linearized Navier–Stokes equations provide an equation-based framework for reduced-order modeling. In particular, bases obtained from the controllability Gramian can be used; these bases are referred to as controllability modes in the literature. When the linearized system is driven by white noise in time and space stochastic forcing, the POD modes obtained from the linearized system are equivalent to the controllability modes. From the work by \cite{cavalieri2022prf}, which used controllability modes in turbulent Couette flow, subsequent examples for the same configuration were studied (\citealt{mccormack2024jfm, maia2024tcfd, maia2025jfm, addisonsmith2025arxiv}). Similar to POD, controllability modes focus only on energetic structures, not necessarily on dynamically important ones. However, from control theory for linearized systems, a modal basis can be built considering the input-output of the system through a balanced truncation \citep{moore1981ieee}, where the modal basis is referred to as balanced  truncation modes. This modal basis in ROMs has been applied successfully to linearized systems (\citealt{willcox2002aiaa, rowley2005ijbc, ilak2008pof, zong2026jfm}), improving the representation of transient growth and harmonic responses. Balanced truncation modes are ranked simultaneously considering controllability and observability, and hence allow a compact representation of the input-output behavior of a linear system. 

ROMs rely on both the modal basis and the equation of motion, even though the latter is not commonly emphasized. In the usual POD-Galerkin setup, the full Navier-Stokes equations are typically used (e.g. \citealt{waleffe1997pof, omurtag1999tcfd, ma2002jfm, noack2003jfm,noack2005jfm, protas2015jfm}); other works rewrite the equations around a known laminar state and solve for the unknown fluctuations (e.g. \citealt{smith2005jfm, ilak2008pof, cavalieri2021prf, cavalieri2022prf,khoo2022jfm, maia2024tcfd, maia2025jfm, zong2026jfm});  and others use a fully data-driven method to model the corresponding equations (e.g. \citealt{sindy_paper, loiseau2018jfm, nakamura2021pof, linot2023jfm, Constante2025jfm}). The relevance of the transition between the laminar solution to the mean turbulent flow over the dynamics has been explored by \cite{noack2003jfm} through an enhanced modal basis using shift modes. A recent work by \cite{zong2026jfm} studied different modal bases in a Galerkin model approach, focusing on the state convergence of these bases in a turbulent Couette flow setup. They found that the POD basis is the best for representing the statistics of turbulence for this configuration. In addition, controllability and balanced truncation modes obtained from linear mean-flow analyses can outperform laminar-flow modal bases in the representation of these statistics. Solving the velocity fluctuations over a mean state is uncommon due to the fact that the equations need to include the effect of the Reynolds stresses, which are typically modeled.

Mean-flow-based analysis has come as a typical framework to understand wall-bounded turbulence \citep{mckeon2010jfm, hwang2010jfm_couette, hwang2010jfm_channelflow}, where the dynamics are decomposed in a linearized operator around the mean state plus an exogenous nonlinear forcing due to the nonlinear interactions. This approach has uncovered significant linear amplification mechanisms, suggesting that much of the energy in turbulent flows is related to the structural properties of the mean shear (e.g. \citealt{delalamo2006jfm, hwang2010jfm_couette, hwang2010jfm_channelflow, mckeon2010jfm, morra2019jfm, von2023aiaa, symon2023prf}). These mechanics are central to the self-sustaining process (SSP) (\citealt{hamilton1995jfm, waleffe1997pof}), a three-stage cycle involving streak formation, streak instability, and the regeneration of streamwise vortices. Linear mechanisms are part of this process through the lift-up effect, where high- and low-speed streaks are generated through the amplification of streamwise velocity perturbations (e.g. \citealt{butler1992pof, delalamo2006jfm, hwang2010jfm_channelflow, hwang2010prf, abreu2020ijhf}), and the Orr mechanism, which describes the transient growth associated with the tilting of vortical structures by the mean shear \citep{jimenez2013pof, encinar2020jfm}.  

The nonlinear forcing in mean-flow-based analysis is typically accounted for either by modifying the linearized operator with an eddy viscosity model \citep{hwang2010jfm_channelflow}, or by treating it as a self-sustaining feedback loop originating from the velocity field \citep{mckeon2010jfm}. Quadratic interactions and the Reynolds stresses are contained in the forcing, and their statistics are spatiotemporally colored \citep{morra2021jfm}. Then, the common assumption of white noise in space and time is inaccurate for representing nonlinear contributions. Several approaches focus either on the availability of statistical data or on approximations to the nonlinear term \citep{zare2017jfm,illingworth2018jfm,towne2020jfm}. ROMs based on Galerkin projection offer a different approach, in which the nonlinear term arises from quadratic interactions among the modal basis functions, without relying on any approximation. A framework where Galerkin projection is combined with the mean-flow-based analysis has not been explored, where the nonlinear interactions are captured explicitly by the Galerkin projection, and the linear operator retains key mechanisms that self-sustain turbulence.

Reduced-order models in wall-bounded turbulence span relatively simple shear flows such as Couette flow (e.g. \citealt{smith2005jfm, cavalieri2022prf, linot2023jfm, zong2026jfm}), sinusoidal flow (Waleffe flow) (e.g. \citealt{waleffe1997pof, moehlis2004njp, cavalieri2021prf}), and pipe flow (e.g. \citealt{constante2024jfm, kaszas2024jfm, Constante2025jfm}). ROMs based on Galerkin projection for turbulent channel flow are not common in the literature, given the change in the boundary conditions compared to turbulent Couette flow and the increased complexity of the simulation \citep{ahmed2021pof}. For the examples available in the literature for ROMs in plane channel flow, POD-Galerkin is a natural choice (e.g. \citealt{podvin1998jfm, omurtag1999tcfd, johansson2006compfluids, mou2023nueng, tsai2025jcp}), but other data-driven approaches are also possible \citep{nakamura2021pof}. Channel flow is a flow that is driven by external pressure; hence, the pressure term is relevant in comparison with Couette flow. With this context, a modification of the ROM based on a linearized equations modal basis is necessary to extend the framework presented in \cite{cavalieri2022prf} for Couette flow to plane channel flow.

Beyond reproducing turbulence statistics, short-time forecasting in ROMs was recently achieved in different works \citep{sindy_paper, loiseau2018jfm,  hijazi2020jcp, callaham2022jfm, linot2023jfm, Constante2025jfm, zong2026jfm}, with the three most recent ones focusing on wall-bounded turbulence. Turbulence, as a chaotic system, displays sensitivity to initial conditions, so only short-term forecasting can be done. Such sensitivity is manifest by the exponential divergence of nearby state-space trajectories,  characterized by the leading Lyapunov exponent $\sim \text{exp}(\lambda_1t)$, which can be calculated for canonical shear flows (e.g. \citealt{keefe1992jfm, inubushi2015physreve, nikitin2018jfm}), leading to a value (in inner units) of $\lambda_1^{+} \approx 0.021$ at the friction Reynolds number $Re_{\tau} = 180$ in turbulent channel flow \citep{nikitin2018jfm} and $\lambda_1^{+} \approx 0.007$ at Reynolds number $Re_{\tau} = 34$ in a minimal flow unit Couette flow \citep{inubushi2015physreve}. The value of the leading Lyapunov exponent is believed to be independent of the boundary conditions and the Reynolds number in near-wall turbulence if the value is scaled in inner units \citep{nikitin2008jfm}. 

In this work, we introduce a ROM based on a Galerkin projection study for turbulent channel flow in a minimal flow unit configuration \citep{jimenez1991jfm}, where a mean-flow-based approach for both the equations and the modal basis is pursued, in contrast with the previous works \citep{cavalieri2022prf, zong2026jfm}. We compare our methodology with the laminar-flow-based ROM from \cite{cavalieri2022prf} and the mean-flow-based basis in a laminar-flow-based ROM \citep{zong2026jfm}. We also compare modal bases from equation-based approaches, such as controllability and balanced truncation modes, with those from data-driven approaches, such as POD modes. First and second-order statistics are evaluated with the different ROMs, assessed against a reference direct numerical simulation (DNS). A modified set of modes is presented to extend the framework present in \cite{cavalieri2022prf} to the plane channel flow setting, where the mean pressure gradient is not negligible as in Couette flow. Finally, the dynamics of the ROMs are compared with short-time forecasting with respect to the reference DNS simulation, considering the highly sensitive conditions of the turbulence through the leading Lyapunov exponent of the turbulent plane channel flow.

The remainder of the paper is organized as follows: in §\ref{sec:modal_bases} the modal bases for plane channel flow are presented; in §\ref{sec:rom} the reduced-order models and the numerical details are described; in §\ref{sec:results} the results and discussion from the statistics and the short-time forecasting are presented for the different ROMs, followed by concluding remarks in §\ref{sec:conclusions}.

\section{Modal bases for plane channel flow}\label{sec:modal_bases}

We consider plane Poiseuille flow for an incompressible fluid with density $\rho$ and kinematic viscosity $\nu$, wherein two infinite planes separated by a wall-normal distance $2h$ remain fixed and a constant bulk velocity $U_b$ is imposed. The domain is defined in Cartesian coordinates by $(x,y,z)$ where $x$ is the streamwise, $y$ is the wall-normal and $z$ is the spanwise direction with dimensions $[0,L_x) \times[-h,h] \times [0,L_z)$, the flow has boundary conditions for the velocity $\boldsymbol{u}|_{y=\pm 1} = (0,0,0)$ in the upper and lower walls and periodic boundary conditions on streamwise and spanwise directions. All variables are dimensionless with respect to $U_b,h$ and $\nu$, where the resulting characteristic number is the Reynolds bulk number $Re_b=U_bh/\nu$. The velocity field is decomposed into a known parallel flow $\boldsymbol{U} = (U(y),0,0)$ and fluctuations around this flow $\boldsymbol{u}^{\prime}(\boldsymbol{x},t)=(u^{\prime}(\boldsymbol{x},t),v^{\prime}(\boldsymbol{x},t),w^{\prime}(\boldsymbol{x},t))$ with $(\boldsymbol{x},t)$, the space and time of the flow field
\begin{equation}
    \boldsymbol{u}(\boldsymbol{x},t) = \boldsymbol{U}(y) + \boldsymbol{u^{\prime}}(\boldsymbol{x},t),
\label{eq:base_fluctuation}
\end{equation}
The velocity fluctuations can be decomposed into a suitable orthonormal modal basis $\boldsymbol{\phi}_i$ with modal coefficients $a_i$ 
\begin{equation}
    \boldsymbol{u} (\boldsymbol{x},t) = \sum_{i=1}^\infty a_i(t)\boldsymbol{\phi}_i(\boldsymbol{x}) \approx\sum_{i=1}^{n} a_i(t)\boldsymbol{\phi}_i(\boldsymbol{x}),
    \label{eq:modal_decomposition}
\end{equation}
which satisfy the boundary conditions and continuity equation, and converge to the velocity fluctuations as $n\rightarrow \infty$. Given the translation invariance of the flow in the streamwise and spanwise directions, we can specify Fourier modes in those directions, where the basis considers the following structure
\begin{equation}
    \boldsymbol{\phi}_i (\boldsymbol{x}) = \boldsymbol{\phi}_{n_xn_z}^{(n)}(\boldsymbol{x}) = \text{exp}\left ( \text{i} \left ( \alpha_on_x x + \beta_o n_z z\right ) \right ) \boldsymbol{\hat{\phi}}_{n_xn_z}^{(n)} (y),
\label{eq:fourier_to_physical}
\end{equation}
where the pair $(n_x,n_z)$ refers to streamwise and spanwise multiples of the fundamental wavenumber $(\alpha_0,\beta_o) = (2\pi/L_x, 2\pi/L_z)$ and the superscript $(n)$ is the $n$-mode associated with that pair $(n_x,n_z)$ ordered by energy (or eigenvalue). In the following, we consider an inner product, which we define in physical space as
\begin{equation}
\langle \boldsymbol{f}, \boldsymbol{g} \rangle = \frac{1}{2hL_xL_z} 
\int_{0}^{L_z} \int_{-h}^{h} \int_{0}^{L_x} \boldsymbol{g}(\boldsymbol{x})^* 
\boldsymbol{f}(\boldsymbol{x})\mathrm{d}x \mathrm{d}y \mathrm{d}z,
\label{eq:inner_product}
\end{equation}
for complex vector fields $\boldsymbol{f}$ and $\boldsymbol{g}$, with $( \cdot )^*$ denoting the Hermitian transpose. In the case of real-valued vectors, we changed the Hermitian transpose to the transpose.

\subsection{POD modes}

POD modes provide a modal basis that optimally represents the kinetic energy in the fluid flow given data from experiments or simulation data \citep{berkooz1993arfm}. Specifically, POD modes are an orthonormal basis that maximizes the projection of the modal basis over the fluid flow, solving the following optimization problem
\begin{equation}
    \underset{\boldsymbol{\phi}}{\text{max}} \frac{\overline{\left| \left\langle \boldsymbol{\phi},\boldsymbol{u^{\prime}} \right\rangle \right|^2 }}{\left\langle \boldsymbol{\phi,\boldsymbol{\phi}} \right\rangle}, 
    \label{eq:max_POD}
\end{equation}
where $\overline{(\cdot)}$ is the ensemble average, or an average over time and periodic directions. The optimization problem can be solved with an eigenvalue problem that considers Fourier modes along streamwise and spanwise directions, given the boundary conditions of the channel \citep{smith2005nd}
\begin{equation}
    \int_{-1}^{1} \overline{\boldsymbol{\hat{u}'}(n_x,y,n_z,t)\boldsymbol{\hat{u}'}^{*}(n_x,y',n_z,t)}
    \boldsymbol{\hat{\phi}}_{n_xn_z}^{(n)} (y') dy' = \lambda_{n_xn_z}^{(n)} \boldsymbol{\hat{\phi}}_{n_xn_z}^{(n)} (y),
\label{eq:eigenvalue_POD}
\end{equation}
such that modes $\boldsymbol{\hat{\phi}}_{n_xn_z}^{(n)} (y)$ can be expressed from Fourier space to physical space using \ref{eq:fourier_to_physical}, and the coefficient $\lambda_{n_xn_z}^{(n)}$ expresses the averaged kinetic energy contained in each mode $(n)$ for a given wavenumber pair $(n_x,n_z)$. Usually solving \ref{eq:eigenvalue_POD} requires a significant amount of snapshots, so we rely on the symmetries of the fluid configuration to data augment the total number of snapshots in the flow field. The symmetries in plane channel flow are typically used to augment the data in the snapshots (e.g. \citealt{omurtag1999tcfd, gibson2008jfm})
\begin{subequations}\label{eq:symmetries}
\begin{align}
I :& (x, y, z, \tilde{u}, \tilde{v}, \tilde{w},p) \mapsto (x, y, z, \tilde{u}, \tilde{v}, \tilde{w},p) \\
R_y :& (x, y, z, \tilde{u}, \tilde{v}, \tilde{w},p) \mapsto (x, -y, z, \tilde{u}, -\tilde{v}, \tilde{w},p) \\
R_z :& (x, y, z, \tilde{u}, \tilde{v}, \tilde{w},p) \mapsto (x, y, -z, \tilde{u}, \tilde{v}, -\tilde{w},p)  \\
R_yR_z :& (x, y, z, \tilde{u}, \tilde{v}, \tilde{w},p) \mapsto (x, -y, -z, \tilde{u}, -\tilde{v}, -\tilde{w},p) 
\end{align}
\end{subequations}
where $p$ is the pressure, $I$ is the identity transformation, $R_y$ is a reflection over $y=0$ plane, $R_z$ is a reflection over $z=0$ and $R_yR_z$ correspond to a 180° rotation over $x-$axis.
\subsection{Linearized equations}

The equations of motion can be decomposed using \ref{eq:base_fluctuation} from a known parallel flow and unknown fluctuations around this flow. The dynamics of the fluctuations can be simplified by taking the Fourier transform in the periodic directions and taking the divergence of the wall-normal velocity $\hat{v}'$ and the wall-normal vorticity $\hat{\eta}'$ as variables. This transformation results in the Orr-Sommerfeld \& Squire equations. Following the same framework as previous works \citep{farrell1993pof,jovanovic2005jfm}, the equations for the fluctuations can be expressed as a linearized dynamical system 
\begin{subequations}
\begin{align}
\frac{\partial \boldsymbol{\hat{q}}}{\partial t} &= \boldsymbol{A}\boldsymbol{\hat{q}} + \boldsymbol{B} \boldsymbol{\hat{f}},\\
\boldsymbol{\hat{u}} &= \boldsymbol{C} \boldsymbol{\hat{q}}.
\end{align} \label{eq:linear_system}
\end{subequations}
where $\boldsymbol{\hat{q}} = \begin{bmatrix} \hat{v}^{\prime}& \hat{\eta}^{\prime} \end{bmatrix}^{T}$, $\boldsymbol{\hat{u}} = \begin{bmatrix} \hat{u}^{\prime}& \hat{v}^{\prime} & \hat{w}^{\prime} \end{bmatrix}^{T}$ is the velocity field in Fourier space along streamwise and spanwise directions and $\boldsymbol{\hat{f}} = \begin{bmatrix} \hat{f}_u& \hat{f}_v & \hat{f}_w \end{bmatrix}^{T}$ are the forcing terms of the linearized system. The operators are defined as
\begin{equation}
    \mathsfbi{A} = \begin{bmatrix} \Delta^{-1}\mathcal{L}_{OS} & 0 \\ - i \beta \mathcal{D}U & \mathcal{L}_{SQ} \end{bmatrix}, \mathsfbi{B} = \begin{bmatrix} - i \alpha \Delta^{-1} \mathcal{D} & - k^2 \Delta^{-1} & - i \beta \Delta^{-1} \mathcal{D} \\ i \beta & 0 & - i \alpha \end{bmatrix}, \mathsfbi{C} = \frac{1}{k^2}\begin{bmatrix} i\alpha \mathcal{D} & -i\beta \\ k^{2}   & 0 \\ i\beta \mathcal{D} & i\alpha \end{bmatrix},
\end{equation}
where the streamwise and spanwise wavenumbers are $(\alpha,\beta) = (\alpha_on_x, \beta_0n_z)$, $\mathcal{D} = d/dy$ is the derivative in wall-normal direction and $\Delta = \mathcal{D}^2-k^2$ is the Laplacian with $k^2=\alpha^2+\beta^2$. The Orr-Sommerfeld \& Squire operators are defined by
\begin{subequations}\label{eq:os-sq}
\begin{align}
\mathcal{L}_{\mathrm{OS}} &= - i \alpha U \Delta + i\alpha \mathcal{D}^{2} U + Re_b^{-1}  \Delta^{2},\\
\mathcal{L}_{\mathrm{SQ}} &= - i \alpha U
+ Re_b^{-1} \Delta .
\end{align}
\end{subequations}
These operators depend on the parallel flow $\boldsymbol{U}$ around which the equations are linearized. Common options are the laminar flow solution $\boldsymbol{U}_0$ or the turbulent mean velocity $\boldsymbol{\bar{U}}$
\begin{equation}
    \boldsymbol{U}_0(y) = U_c\left(h^2-y^2,0,0 \right), \quad \quad \quad \quad \boldsymbol{\bar{U}}(y) = \left(\bar{U}(y),0,0 \right),
    \label{eq:def_U}
\end{equation}
where the centerline velocity is $U_c = 3/2$, due to the condition of the bulk velocity $U_b=1$.
\subsection{Controllability modes}

Controllability modes identify the directions in state space $\boldsymbol{\hat{q}}$ that are most easily reached by external inputs $\boldsymbol{\hat{f}}$. For the linearized Navier–Stokes equations driven by white-in-time and space stochastic forcing, these modes correspond to the most energetic structures of the system response, i.e. the POD modes of the forced dynamics. They are obtained from the infinite-time controllability Gramian,
\begin{equation}
\mathsfbi{W}_c^q = \int_{0}^{\infty} e^{\boldsymbol{A} t}\mathsfbi{B}\mathsfbi{B}^{*} e^{\boldsymbol{A}^{*} t}dt, \label{eq:control_gramian}
\end{equation}
Usually, the direct computation of the Gramian is numerically intractable, so it is numerically better to solve the Lyapunov equations instead
\begin{equation}
\mathsfbi{A}\mathsfbi{W}_c^q + \mathsfbi{W}_c^q\mathsfbi{A}^* + \mathsfbi{B}\mathsfbi{B}^*=0,
\label{eq:lyapunov_cntrl}
\end{equation}
With the controllability Gramian, the modal bases can be obtained solving the following eigenvalue problem for different wavenumber pairs $(n_x,n_z)$
\begin{equation}
    \mathsfbi{W}_c^q \boldsymbol{\hat{\Phi}}^q = \boldsymbol{\hat{\Phi}}^q \Lambda,
    \label{eq:eigenvalue_cntrl}
\end{equation}
The eigenvector can be transformed to primitive variables with $\boldsymbol{\hat{\Phi}} = \boldsymbol{C}\boldsymbol{\hat{\Phi}}^q$, then $\boldsymbol{\hat{\Phi}} = [\boldsymbol{\hat{\phi}}^{(1)}\boldsymbol{\hat{\phi}}^{(2)}\cdots\boldsymbol{\hat{\phi}}^{(n)}]$ and eigenvalues are ordered by energy $\Lambda = [\lambda^{(1)}\lambda^{(2)}\cdots\lambda^{(n)}]$ where we dropped the explicit dependence to the wavenumber pairs $(n_x,n_z)$. The resulting modes are optimal in a 2-norm $\left\| \boldsymbol{q} \right\|_2^2$, but it is often necessary to consider an appropriate energy $Q-$norm for the state vector  $\left\| \boldsymbol{q} \right\|_Q^2 = \boldsymbol{q}^*\mathsfbi{Q}\boldsymbol{q}$. In particular, if we have an appropriate weight matrix, the following Q-norm matrix can be decomposed into a Cholesky form $\mathsfbi{Q}  = \mathsfbi{F}^* \mathsfbi{F}$ \citep{herrmann2018ijhmt, herrmann2023jfm}. Then the linear system is transformed accordingly with $\boldsymbol{\tilde{q}} = \mathsfbi{F}\boldsymbol{q}$, and the dynamics to $\mathsfbi{\tilde{A}} = \mathsfbi{F}\mathsfbi{A}\mathsfbi{F}^{-1}$, $\mathsfbi{\tilde{B}} = \mathsfbi{F}\mathsfbi{B}$, $\mathsfbi{\tilde{C}} = \mathsfbi{C}\mathsfbi{F}^{-1}$. This transformation results in modes that are optimal in an energetic Q-norm sense.

\subsection{Balanced truncation modes}

Balanced truncation yields a modal basis that orders states according to their joint controllability and observability \citep{moore1981ieee, willcox2002aiaa}. In other words, it identifies states that are both easily excited by inputs and strongly reflected in the outputs. This is achieved through the use of the controllability and observability Gramians, where the latter is defined by
\begin{equation}
    \boldsymbol{W}_o^q = \int_{0}^{\infty} e^{\boldsymbol{A}^{*}  t}\mathsfbi{C}^{*} \mathsfbi{C} e^{\boldsymbol{A} t}\, dt, \label{eq:observability_gramian} 
\end{equation}
Similar to controllability Gramian, the infinite-time observability Gramian is numerically intractable, so it is often solved directly with the Lyapunov equations
\begin{equation}
\mathsfbi{A}^*\mathsfbi{W}_o^q + \mathsfbi{W}_o^q\mathsfbi{A} + \mathsfbi{C}^*\mathsfbi{C}=0,
\label{eq:lyapunov_obs}
\end{equation}
The balanced transformation seeks a coordinate transformation matrix $\mathsfbi{T}^q$ where the Gramians are equal and diagonal \citep{brunton2022data}. This leads to
\begin{align}
\tilde{\mathsfbi{W}_c^q}\tilde{\mathsfbi{W}_o^q} =& \left( \mathsfbi{T}^q \right)^{-1}\mathsfbi{W}_c^q \left( \mathsfbi{T}^q \right)^{-*}\left( \mathsfbi{T}^q \right)^*\mathsfbi{W}_o^q  \mathsfbi{T}^q \\
=& \left( \mathsfbi{T}^q \right)^{-1}\mathsfbi{W}_c^q \mathsfbi{W}_o^q \mathsfbi{T}^q = \Lambda,
\end{align}
with $(\cdot)^{-*}$ corresponding to the inverse of the Hermitian transpose. Then the matrix $\mathsfbi{T}^q$ can be found by solving the following eigenvalue problem
\begin{equation}
    \mathsfbi{W}_c^q \mathsfbi{W}_o^q \mathsfbi{T}^q = \mathsfbi{T}^q \Lambda,
    \label{eq:eigenvalue_balan}
\end{equation}
The balanced  transformation can be expressed in terms of primitive variables with $\mathsfbi{T} = \mathsfbi{C}\mathsfbi{T}^q$ where the direct modes are defined by $\mathsfbi{T} = [\boldsymbol{\hat{\phi}}^{(1)}\boldsymbol{\hat{\phi}}^{(2)}\cdots\boldsymbol{\hat{\phi}}^{(n)}]$, the adjoint modes correspond to $\mathsfbi{T}^{-1} = \mathsfbi{S} =[\boldsymbol{\hat{\psi}}^{(1)}\boldsymbol{\hat{\psi}}^{(2)}\cdots\boldsymbol{\hat{\psi}}^{(n)}]$ and the Hankel singular values that relate with the dynamical relevance of each mode $\Sigma = [\sigma^{(1)}\sigma^{(2)}\cdots\sigma^{(n)}]$. The set of modes is biorthonormal and not orthonormal as POD and controllability modes, satisfying then $\langle \boldsymbol{\hat{\phi}}^{(i)}, \boldsymbol{\hat{\psi}}^{(j)} \rangle = \delta_{ij}$

\subsection{Galerkin (Petrov-) projection}

Reduced-order models seek a modal decomposition in \ref{eq:modal_decomposition} that significantly reduces the number of degrees of freedom from the full model. This decomposition depends, as we described, on the modal basis, but also on the equation that we choose to model the dynamics. The ROMs can be described generally with the equation
\begin{equation}
    \frac{d a_i}{dt} = \left\langle \boldsymbol{\psi}_i ,\mathcal{F} \left ( \boldsymbol{\Phi} \boldsymbol{a}\right) \right\rangle,
    \label{eq:ROM_general}
\end{equation}
where the modal coefficients are $\boldsymbol{a} =[a_1,a_2,a_3,\cdots, a_n]^T$ and the modal bases are $\boldsymbol{\Phi} =[\boldsymbol{\phi}_1,\boldsymbol{\phi}_2,\boldsymbol{\phi}_3,\cdots, \boldsymbol{\phi}_n]$. The ROM depends on the modal basis $\boldsymbol{\Phi}$ chosen in \ref{eq:modal_decomposition} and on the dynamics $\mathcal{F}(\cdot)$ which describes the fluid flow. The projection (Petrov-Galerkin) used in \ref{eq:ROM_general} considers a biorthonormal basis $\langle \boldsymbol{\phi}_{i}, \boldsymbol{\psi}_{j} \rangle = \delta_{ij}$, but in particular, if we have an orthonormal basis, then we use a Galerkin projection with $\boldsymbol{\Psi}=\boldsymbol{\Phi}$. An appropriate choice of both the modal basis and the dynamical model is crucial for the performance of the ROM, as discussed in the following sections.

\section{Mean-flow-based reduced-order models}\label{sec:rom}

Within the framework of mean-flow-based analysis, the equations of motion are described around a mean state $\boldsymbol{\bar{U}}$ commonly averaged over time and homogeneous directions. The total velocity and pressure fields are decomposed as $\boldsymbol{u} = \boldsymbol{\bar{U}} + \boldsymbol{u}'$ and $p = \bar{P} + p'$, where $\boldsymbol{u}'$ and $p'$ denote the fluctuations about the mean. Substituting this decomposition in the Navier-Stokes equation and subtracting the equation of the mean flow results in the following equations for the fluctuation field $\boldsymbol{u}'$ 
\begin{subequations}\label{eq:nse_equations}
\begin{align}
\frac{\partial \boldsymbol{u}'}{\partial t} + \left ( \boldsymbol{u}' \cdot \nabla \right ) \boldsymbol{\bar{U}} + \left ( \boldsymbol{\bar{U}} \cdot \nabla \right ) \boldsymbol{u}' + \nabla p' - \frac{1}{Re_b} \nabla^2 \boldsymbol{u}' =  - \left ( \boldsymbol{u}' \cdot \nabla \right ) \boldsymbol{u}' + \overline{\left ( \boldsymbol{u}' \cdot \nabla \right ) \boldsymbol{u}'}, \label{eq:momentum} \\
\boldsymbol{\nabla} \cdot \boldsymbol{u}' = 0.
\label{eq:continuity}
\end{align}
\end{subequations}

The equations of momentum in \ref{eq:momentum} are forced by a zero mean forcing term $\boldsymbol{G}(t) = G(t) \boldsymbol{e}_x$ that keeps the mass flux constant in the channel.
As the mean state is not a solution of the Navier-Stokes equations, the Reynolds stresses $\overline{\left ( \boldsymbol{u}' \cdot \nabla \right ) \boldsymbol{u}'}$ appear in the equations of motion. In the left-hand side of the equation, the linear operators are defined around a mean state and not around a laminar state as previous ROMs \citep{cavalieri2022prf, maia2025jfm,zong2026jfm}. Both differences change the physical meaning of the velocity fluctuations and how the linear mechanisms that sustain the turbulence are captured in the equations

\subsection{Plane channel flow setup}

In plane channel flow we consider the laminar base state as $\boldsymbol{U}_0 = U_0(y) \boldsymbol{e}_x$ and the mean state is averaged in time and spatial directions resulting in $\boldsymbol{\bar{U}} = \bar{U}(y)\boldsymbol{e}_x$ as in Section \ref{sec:modal_bases}. This mean state is modeled by the Reynolds-averaged Navier-Stokes equations, which for the considerations in the base flow are the following
\begin{equation}\label{eq:base_state_eqs}
    -\nabla \bar{P} + \frac{1}{Re_b} \frac{d^2 \bar{U}}{dy^2} \boldsymbol{e}_x - \overline{\left ( \boldsymbol{u}' \cdot \nabla \right ) \boldsymbol{u}'} = 0,
\end{equation}
The projection of the equations in \ref{eq:nse_equations} can be done using any modal decomposition from \ref{sec:modal_bases} with the general projection \ref{eq:ROM_general}. The projection of the left-hand side (linear terms) in \ref{eq:momentum} is direct as other works \citep{cavalieri2022prf, maia2025jfm}, where only the physical meaning of the terms changes. The projection of the right-hand side (non-linear terms) in the mean-flow-based framework requires further work, given the influence of the Reynolds stresses, where they can be reworked using the base state equation \ref{eq:base_state_eqs}
\begin{equation}
    \begin{aligned}
            \left\langle  \overline{\left ( \boldsymbol{u}' \cdot \nabla \right ) \boldsymbol{u}'} ,\boldsymbol{\psi}_i \right\rangle &= \left\langle -\nabla \bar{P},\boldsymbol{\psi}_i \right\rangle +  \frac{1}{Re_b}\left\langle \frac{d^2 \bar{U}}{dy^2} \boldsymbol{e}_x,\boldsymbol{\psi}_i \right\rangle, \\
            &= \int_{\Gamma} -\bar{P}(\boldsymbol{\psi}_i\cdot\boldsymbol{n})d\Gamma - \left\langle -\bar{P},\nabla \cdot \boldsymbol{\psi}_i \right\rangle + \frac{1}{Re_b}\left\langle \frac{d^2 \bar{U}}{dy^2} \boldsymbol{e}_x,\boldsymbol{\psi}_i \right\rangle, 
    \end{aligned}
    \label{eq:pressure_baseflow}
\end{equation}
Using the divergence theorem, the projection over the mean pressure can be split into two different terms as shown in \ref{eq:pressure_baseflow}. This equation states that the projection of the Reynolds stresses involves three different terms: the first is a boundary integral in $\Gamma$ which for plane channel flow is not zero given that the base state is driven by the mean pressure gradient, the second involves a projection of the mean pressure over the divergence of the modes which is zero given the divergence-free condition of the modes in \ref{eq:continuity} and the third involves a projection of the Laplacian of the mean flow over the modal basis.

\subsection{The need for a modified basis: Modified Stokes modes}

The projection of the Reynolds stresses involves projecting the mean pressure gradient onto the modal bases. This is in contrast with Couette flow, where the mean pressure gradient can be neglected because the boundary conditions drive the flow. Plane channel flow is pressure-driven; thus, the pressure gradient sustains the flow.

\begin{figure}
  \centerline{\includegraphics[width=1\textwidth]{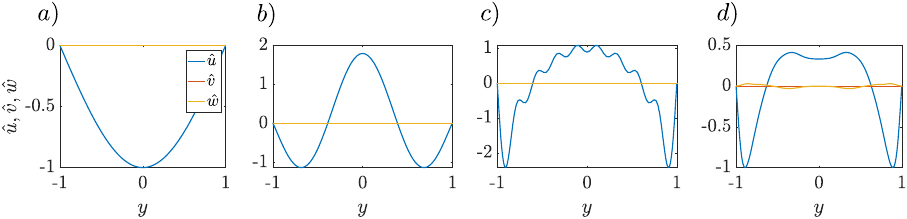}}% Images in 100% size
  \caption{Modes $\boldsymbol{\psi}_{00}^{(1)}$ used in the construction of the modal basis, plotted in Fourier space. (a) First Stokes mode, (b) First modified Stokes mode for $n=9$ total modes, (c) First modified Stokes mode for $n=36$ total modes, (d) First POD mode. The ranking of the modes is through eigenvalues of the same wavenumber pair $(n_x,n_z)=(0,0)$}
\label{fig:stokes_modes}
\end{figure}

In plane channel flow the mean pressure gradient can be assumed constant with a variable mass flux in the channel, or a forcing $\boldsymbol{G}(t)$ it can be added to the mean pressure gradient to satisfy a constant mass flux in the channel. As the mean flow has only a streamwise component, it can be assumed that the forcing has only a streamwise component. The projection of the mean pressure gradient can be added as an additional time-dependent term to the ROM that modifies the mean pressure gradient at each step, or it can be imposed by modifying the modal bases to have a null projection as shown below
\begin{equation}
    \left\langle  \boldsymbol{e}_x,\boldsymbol{\psi}_i \right\rangle =  \int_{-1}^{1} \boldsymbol{\hat{\psi}}_i(y')\boldsymbol{e}_x dy' = 0,
\label{eq:stokes_mod_cond}
\end{equation}
Since this condition refers to the mean flow, only modes with $\alpha=\beta=0$ are relevant. In data-driven modal bases, such as POD, this condition is satisfied because the simulation data can be forced to satisfy a mass-flux condition. However, for modes from linearized equations, this condition is not necessarily satisfied. As the operators presented in \ref{eq:linear_system} are singular for this wavenumber pair, works rely on eigenfunctions of the viscous operators i.e. Stokes modes (e.g. \citealt{waleffe1997pof, cavalieri2022prf}). The condition to enforce in the Stokes modes is to subtract the mean value in the streamwise direction, which the following projections can define
\begin{equation}
    \mathcal{P}_\parallel  (\boldsymbol{\hat{\psi}}_i) = \frac{1}{2}\int \boldsymbol{\hat{\psi}}_i(y') \boldsymbol{e}_x dy', \quad \quad \quad \quad \mathcal{P}_\bot  (\boldsymbol{\hat{\psi}}_i) = \boldsymbol{\hat{\psi}}_i -\mathcal{P}_\parallel(\boldsymbol{\hat{\psi}}_i), 
\end{equation}
The modified Stokes basis can be defined using the orthogonal projector over the Stokes basis $\mathcal{P}_\bot  (\boldsymbol{\hat{\psi}}_i)$. The resulting modal basis is not orthogonal but the orthogonality can be recovered using a Gram-Schmidt process or through the left singular vectors doing a singular value decomposition (SVD) over the non-orthogonal basis. For the remainder of this work, the latter approach is adopted. The resulting orthogonal modal basis is $\tilde{\boldsymbol{\Psi}}  = [\boldsymbol{\tilde{\psi}}_{1}\boldsymbol{\tilde{\psi}}_{2}\cdots\boldsymbol{\tilde{\psi}}_{n}]$, which satisfies $\langle \boldsymbol{\tilde{\psi}}_{i}, \boldsymbol{\tilde{\psi}}_{j} \rangle = \delta_{ij}$. An example of the structures we recover using the left singular vectors of an SVD with this process can be seen in Figure \ref{fig:stokes_modes}. In the case of balanced truncation modes, we rely on the same modal bases $\boldsymbol{\tilde{\Phi}} = \boldsymbol{\tilde{\Psi}}$

\subsection{Galerkin (Petrov-) projection}

We can use a Petrov-Galerkin projection in \ref{eq:momentum}, using the general equation in \ref{eq:ROM_general} for any suitable modal basis. The resulting projection gives
\begin{equation}
  \frac{\mathrm{d}a_i}{\mathrm{d}t} 
= \sum_{j=1}^n L_{ij} a_j
+ \sum_{j=1}^n \tilde{L}_{ij} a_j
+ \sum_{k=1}^n \sum_{j=1}^n Q_{ijk} a_ja_k + F_i,
  \label{eq:ROM_eq}
\end{equation}
where the projection on the pressure term is zero if we use modified Stokes modes that satisfy \ref{eq:stokes_mod_cond}. The operators are defined by
\begin{subeqnarray}
    L_{ij} &=& Re_b^{-1} \langle \nabla^2 \boldsymbol{\phi}_j , \boldsymbol{\psi}_i \rangle, \\
    \tilde{L}_{ij} &=& - \langle  (\boldsymbol{\phi}_j \cdot \nabla) \boldsymbol{U} + (\boldsymbol{U} \cdot \nabla) \boldsymbol{\phi}_j , \boldsymbol{\psi}_i \rangle, \\
    Q_{ijk} &=& - \langle (\boldsymbol{\phi}_j \cdot \nabla )\boldsymbol{\phi}_k , \boldsymbol{\psi}_i \rangle, \\
    F_i &=&  Re_b^{-1}\left\langle \frac{d^2 \bar{U}}{dy^2} \boldsymbol{e}_x, \boldsymbol{\psi}_i  \right\rangle,
    \label{eq:ROM_op}
\end{subeqnarray}
These operators are similar to the known framework in the literature (e.g. \citealt{cavalieri2022prf, maia2024tcfd, zong2026jfm}). In the case of laminar-flow-based operators, the projection of the $F_i$ is zero by construction, but in the mean-flow-based approach, this quantity takes into account the projection of the Reynolds stresses. We consider this as a general model for both laminar-flow-based and mean-flow-based models, where the model is changed with respect to the definition of base flow used $\boldsymbol{U}(y)$ and in the inclusion of the $F_i$ term. The different options we covered for the ROMs are explained in Table \ref{tab:roms_config} 

\begin{table}
  \begin{center}
    \def~{\hphantom{0}}
  \begin{tabular}{ccccc}
Case & $\phi_i/\psi_i$ & $U$ (Modal bases) & $\mathcal{F}$ (Model) & $F_i$ (Reynolds stresses)\\[8pt]
    C-LL     & Controllability modes     & $U_0(y)$  & Laminar-flow-based & $\times$ \\
    C-TL   & Controllability modes     & $\overline{U}(y)$  & Laminar-flow-based & $\times$  \\
    C-TT   & Controllability modes     & $\overline{U}(y)$ & Mean-flow-based &   \checkmark   \\
    BT-LL  & Balanced truncation modes & $U_0(y)$ & Laminar-flow-based & $\times$    \\
    BT-TL  & Balanced truncation modes & $\overline{U}(y)$ & Laminar-flow-based  & $\times$  \\
    BT-TT  & Balanced truncation modes & $\overline{U}(y)$ & Mean-flow-based & \checkmark  \\
    POD-T     & POD modes of turbulent state & Not applicable & Mean-flow-based & \checkmark  \\
  \end{tabular}
  \caption{List of different ROMs compared in this work}
  \label{tab:roms_config}
  \end{center}
\end{table}

% An additional relation can be made to build an equation energy as \cite{maia2025jfm}, where the fluctuation kinetic energy is

% \begin{equation}
%     \frac{dE}{dt} = \frac{d}{dt} \left ( \frac{1}{2} \langle  \boldsymbol{u}, \boldsymbol{u} \rangle \right ) = \frac{d}{dt} \left ( \frac{1}{2} \sum_i^n \sum_j^n a_i a_j \langle  \boldsymbol{\phi}_i, \boldsymbol{\phi}_j \rangle \right ) = \frac{d}{dt} \left( \frac{1}{2}\boldsymbol{a}^T \boldsymbol{M}\boldsymbol{a}\right )
% \end{equation}

\subsection{Numerical implementation}

We define a mesh $(N_x,N_y,N_z)=(16,101,32)$ using equidistant points in the streamwise and spanwise directions. The wall-normal direction is discretized with Chebyshev polynomials, whereas the homogeneous directions are discretized with Fourier modes. A reference DNS simulation is obtained using the Channelflow spectral code \citep{gibson2008jfm,gibson2014chflow} with a factor of $3/2$ in $N_x$ and $N_z$ for de-aliasing, and statistics are also collected after a transient of 100 time units. We choose the dimensions of the channel equal to $(L_x,L_y,L_z)=(1.81,2,0.92)$ with $Re_b =2800$ similarly to a minimal flow unit \citep{jimenez1991jfm}, where the resulting friction Reynolds $Re_{\tau} = h/\delta_v \approx 185$ considering $\delta_\nu = \nu/u_{\tau}$ is the viscous length scale and $u_{\tau}$ is the friction velocity. Quantities with superscript $+$ denote inner (or viscous) units which are scaled in length by $\delta_\nu$, in velocity with $u_{\tau}$, and in time with $\nu/u_{\tau}^2$

For the modal basis, we consider $\left| n_x \right| \leq 2, \left| n_z \right| \leq1$ Fourier wavenumber pairs. In the case of POD modes, we consider the DNS reference simulation to build the eigenvalue problem in \ref{eq:eigenvalue_POD}, using the symmetries in \ref{eq:symmetries} for data augmentation, after the same transient of 100 time units. For the framework described in \ref{eq:linear_system}, the controllability and observability Gramians are obtained by solving the Lyapunov equations described in \ref{eq:lyapunov_cntrl}, \ref{eq:lyapunov_obs}, then controllability modes are obtained by solving the eigenvalue problem in \ref{eq:eigenvalue_cntrl} and balanced modes solving problem \ref{eq:eigenvalue_balan}. For all three eigenvalue problems described, we order the resulting eigenvectors by eigenvalue (or Hankel singular value in the balanced  truncation case) and truncate by $n_y = 9,18,27,36$. We decompose for every modal basis the real and imaginary part of the mode in \ref{eq:fourier_to_physical}, which gives two sets of modes separated by a phase shift of $\pi/2$. Then, modes with $(-n_x,-n_z)$ are dropped in the basis, as they are complex conjugates of $(n_x,n_z)$, so only positive $n_x$ is considered. For $(n_x,n_z)=(0,0)$ in the equation-based approach, we choose the modified Stokes modes described in \ref{eq:stokes_mod_cond}. The total amount of modes for these configurations is then ranges over $n=135,270,405,540$ modes

The ROM defined in \ref{eq:ROM_eq} is run using an initial condition $a_i \in (0,0.5)$. The system is integrated up to 3000 non-dimensional time units ($tU_b/h$) with a 0.5 step using the \texttt{ode45} function in MATLAB. Statistics are collected after the first 1000 time units. The base state where the fluctuations are defined is given in \ref{eq:def_U}, where the average streamwise velocity is obtained through the DNS.

The error between different truncations $n_y$ is quantified by the $L^2$ norm of the difference for the chosen value to compare \citep{maia2024tcfd}. For instance, the error in the RMS values of the streamwise component of the velocity fluctuation is
\begin{equation}
    \varepsilon_{(u_{rms})} = \frac{\left\| (u_{rms})_{\text{(DNS)}} - (u_{rms})_{\text{(ROM)}} \right\|_2}{\left\| (u_{rms})_{\text{(DNS)}} \right\|_2},
    \label{eq:err_stats}
\end{equation}
The $L^2$ norm is equivalent to the inner product as $\left\| \boldsymbol{u} \right\|_2 = \sqrt{\langle \boldsymbol{u}, \boldsymbol{u} \rangle}$, and the RMS values are defined, for example, by $u'_{rms} = \sqrt{(\overline{u'u'})}$. 

\section{Results \& Discussion}\label{sec:results}

We compare the models in terms of convergence of statistical quantities, as usual in the literature, and we also expand the comparison to short-time forecasting, focusing on the dynamical agreement of the ROMs with respect to DNS. 

\subsection{Turbulence statistics}

\begin{figure}
\centerline{\includegraphics[width=0.8\textwidth]{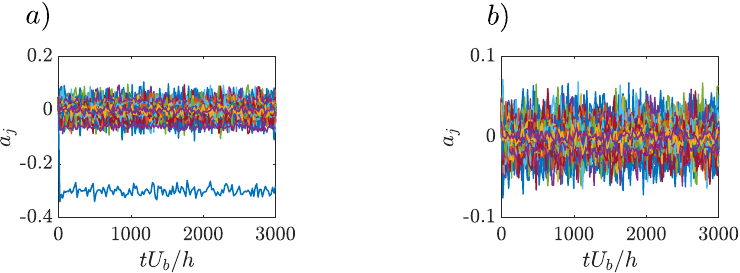}}% Images in 100% size
  \caption{Modal coefficients $a_j(t)$ time series for (a) C-TL (b) C-TT. Both models use $n_y=18$ ($n=270$).}
\label{fig:modal_timeseries}
\end{figure}

The time series of the modal coefficients for two different models using the same modal basis (controllability modes) are shown in Figure \ref{fig:modal_timeseries}. As the model $\mathcal{F}$ changes between the two ROMs, the dynamics of the modal coefficients differ. For instance, in the mean-flow-based ROM, all the mode coefficients tend to have zero mean. In contrast, the laminar-flow-based model does not necessarily satisfy this, as mean-flow modes have non-zero time average, leading to a mean flow different from the laminar solution. Both models exhibit chaotic behavior, as shown in previous works \citep{cavalieri2022prf, maia2025jfm}. The chaotic behavior is not necessarily observed in every ROM simulated in this work, where models based on balanced truncation (BT-LL, BT-TL, and BT-TT), on the other hand, suffer from blow-ups as reported in \cite{zong2026jfm} when a low number of eigenvectors $n_y$ are used in the modal bases. The initial condition was changed from $a_i \in (0, 0.1)$ used in \cite{cavalieri2022prf} to $a_i \in (0, 0.5)$ to prevent re-laminarization of the models. The blow-ups can be further explained through the nonlinear term $Q_{ijk}$, which is known to be a conservative term \citep{loiseau2018jfm, callaham2022jfm}, but when balanced truncation modes are used, this property is not necessarily satisfied because the modal bases are not orthogonal $ \langle  \boldsymbol{\psi}_j, \boldsymbol{\psi}_i \rangle \neq \delta_{ij} $

\begin{figure}
\centerline{\includegraphics[width=1\textwidth]{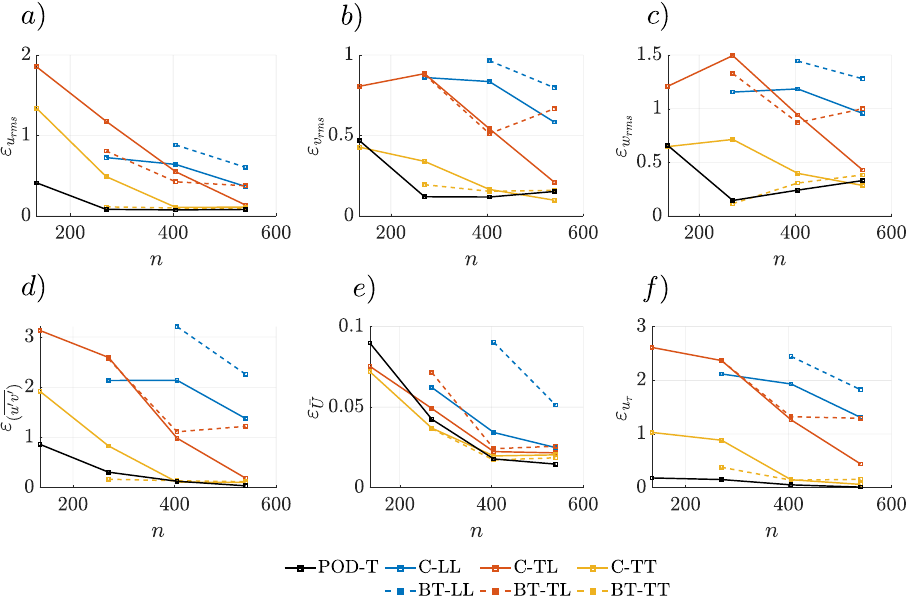}}% Images in 100% size
  \caption{Errors of the first and second order statistics defined in \ref{eq:err_stats}. (a) Streamwise component values, (b) Wall-normal component values, (c) Spanwise component values , (d) $\overline{u'v'}$ stress component (e) Mean flow $\bar{U}$, f) Friction velocity $u_{\tau}$. The models missing in the plots show blowing-up behavior}
\label{fig:error_rms_umean}
\end{figure}

The convergence of different statistics is shown in Figure \ref{fig:error_rms_umean}. We compare different modal bases and models, focusing on the changes in convergence between laminar-flow-based ROMs and mean-flow-based ROMs. For the comparison of statistics of the rms values, we show in Figure \ref{fig:error_rms_umean}a,b,c, the best representation of the statistics is given by the POD modal basis, as recently shown in \cite{zong2026jfm} in a Couette flow setup. The use of a modal basis based on a mean flow provides a better representation of RMS values than relying on a basis from a laminar state. However, the choice of model is also key here, maintaining the same modal basis while changing the model can significantly improve the statistics, yielding results comparable to those of the POD model. We also observed that balanced truncation modes do not provide a significant improvement over controllability modes, even when the projected model changes.

The convergence of the mean flow and the friction velocity is shown in Figure \ref{fig:error_rms_umean}d,e,f, where all the cases converge to a similar error of the mean flow, with the exception of the BT-LL case. The convergence of the friction velocity gives an approximation of the convergence of the friction Reynolds number, where mean-flow-based ROMs achieve a closer value to the friction velocity than laminar-flow-based ROMs, given that this is an input of the model through the decomposition of the velocity in \ref{eq:base_fluctuation}. Mean-flow-based ROMs verify that the recovered friction velocity and $\overline{u'v'}$ component of the Reynolds stresses is only achieved when sufficient modes are given ($n>400$), Laminar-flow-based ROMs do not recover these values with exception of C-TL when $n=540$ modes are used. This implies that laminar-flow-based models can achieve comparable results to mean-flow-based models but need more modes for an equivalent results.

\begin{figure}
\centerline{\includegraphics[width=0.95\textwidth]{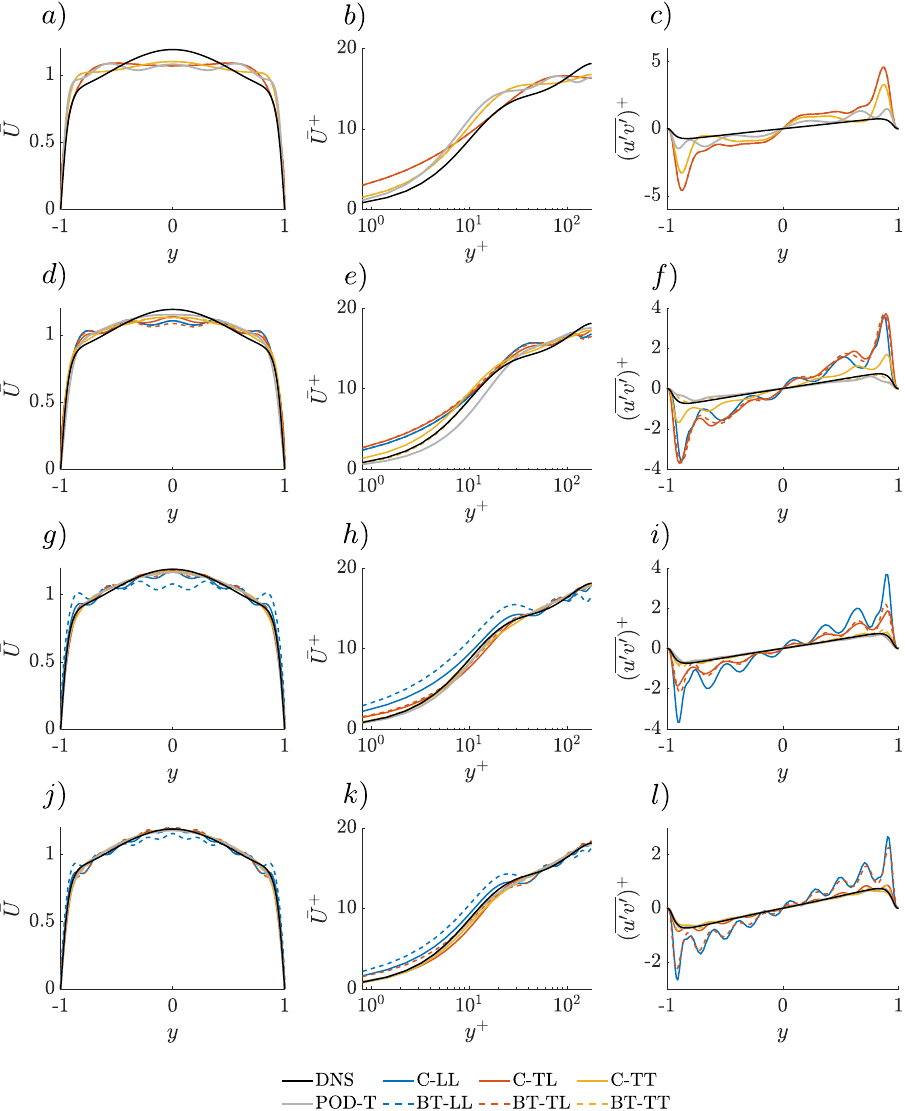}}% Images in 100% size
  \caption{Statistics of the models $\overline{\boldsymbol{U}}$ in outer units (a-d-g-j), in inner units (b-e-h-k), $\overline{u'v'}$ in (c-f-i-l).  Models using $\left| n_x \right| \leq 2, \left| n_z \right| \leq 1$ and (a-c) $n_y = 9$, (d-f) $n_y=18$, (g-i) $n_y=27$, (j-l) $n_y=36$, equivalent to $n=135, 270, 405, 540$. The models missing in the plots show blowing-up behavior. For $\overline{u'v'}$ statistics, the BT-LL is not shown due to high deviation with respect to other models}
\label{fig:umean_comparison}
\end{figure}

\begin{figure}
  \centerline{\includegraphics[width=0.95\textwidth]{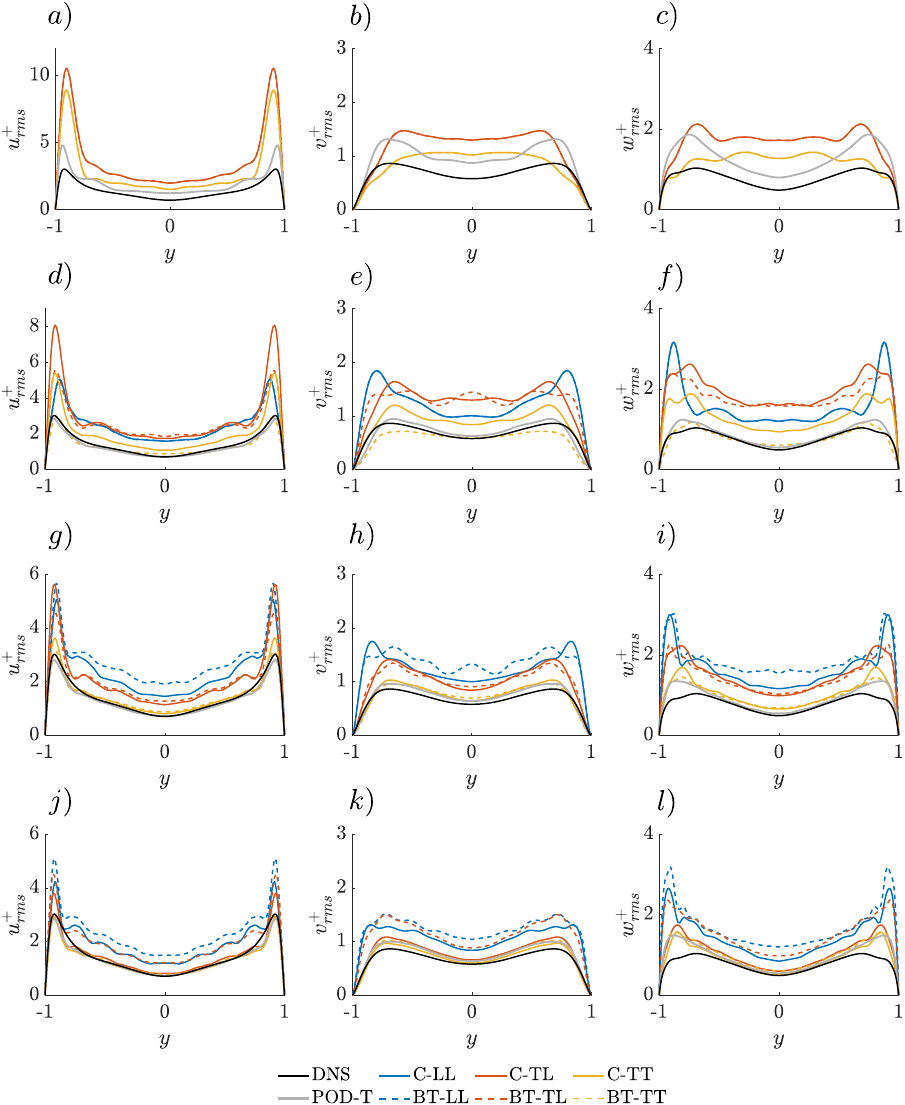}}% Images in 100% size
  \caption{Statistics of the models $u_{rms}$ in (a-d-g-j), $v_{rms}$ in (b-e-h-k), $w_{rms}$ in (c-f-i-l).  Models using $\left| n_x \right| \leq 2, \left| n_z \right| \leq 1$ and (a-c) $n_y = 9$, (d-f) $n_y=18$, (g-i) $n_y=27$, (j-l) $n_y=36$, equivalent to $n=135, 270, 405, 540$. The models missing in the plots show blowing-up behavior}
\label{fig:rms_comparison}
\end{figure}

The behavior of the error in the different plots of Figure \ref{fig:error_rms_umean} is non-monotonic as previous works on ROMs \citep{cavalieri2022prf, addisonsmith2025arxiv, zong2026jfm} and as has also been reported in DNS with coarse grids \citep{meyers2007pof, rasam2011jot}. The values of POD-T, C-TL, C-TT, and BT-TT converge to similar statistics when a sufficient number of modes is provided, as appears in Figure \ref{fig:error_rms_umean}; the other models do not converge to the same statistics due to the limited number of modes and both modal bases and equations used as discussed. The values can be compared with the reported values in \cite{zong2026jfm} for $n_y=10$, equivalently $n=150$ modes (see his Table 4), and with our results for $n=135$ modes. The errors for the second-order statistics (RMS) are higher than in his study because they use a lower friction Reynolds number ($Re_{\tau} \approx 34$). The effect of the equations is non-negligible, as can be seen in Figure \ref{fig:error_rms_umean}, where, for a low amount of modes, the mean-flow-based approach can achieve statistics similar to POD-Galerkin at least in the wall-normal and spanwise directions.

A further visualization in physical space of the statistics is given in Figures \ref{fig:umean_comparison} and \ref{fig:rms_comparison}, where in Figure \ref{fig:umean_comparison} the convergence of the mean flow and the $\overline{u'v'}$ are shown. The use of modified Stokes ensures that the mean flow obtained has the same bulk velocity as the one from DNS even though the values are different between the ROM and the DNS as presented in Figure \ref{fig:umean_comparison}a,d,g,j. However, the values near wall in Figure \ref{fig:umean_comparison}b,e,h,k demonstrate that accuracy on the mean velocity representations does not imply good representation in the near wall velocity. Even though ROMs achieve low error as presented in Figure \ref{fig:error_rms_umean}e, laminar-flow-based ROMs (specifically C-LL and BT-LL) do not show a good representation of the near wall values. The inclusion of mean-flow-based modal bases (C-TL and BT-TL) improves the results, but models that use both mean-flow-based equations and modal bases (C-TT, BT-TT) can achieve results comparable to POD-T, as indicated by Figure \ref{fig:error_rms_umean}f. For the visualization of the RMS values, we present Figure \ref{fig:rms_comparison}, where the three components of the RMS values are present. The models, as discussed, give reasonable accuracy in a further reduced space compared with the DNS, with closer agreement for the mean-flow-based models. 

Overall, these results highlight that one may achieve better low-order representation of the turbulence dynamics by considering equations taken around the mean flow. There is a significant body of literature showing that mean-flow based modes offer a good representation of coherent structures in turbulence (see \citealt{mckeon2017jfm} and references therein), and here we extend this view to the non-linear dynamics, as the TT models, with mean-flow based modes and dynamics show better agreement with the reference statistics.

\subsection{Short-time forecasting}

A comparison of the short-time forecasting between the ROMs and the DNS is studied, where we define the following initial condition for the ROMs
\begin{equation}
    a_i(t=0) =\langle \boldsymbol{u'}, \boldsymbol{\phi}_i \rangle,
    \label{eq:projection_velocity}
\end{equation}
where the velocity field $\boldsymbol{u'}$ is obtained from the DNS. The projection depends on the modal basis chosen $\boldsymbol{\phi}_i$ and the definition of the modal coefficients used in the ROM, presented in \ref{eq:base_fluctuation} and \ref{eq:modal_decomposition}.

To study the behavior of the models in the short-time forecasting, we define the following approach: from the DNS, we randomly choose different snapshots and then project them with \ref{eq:projection_velocity} in the ROM to run a parallel simulation in a short-time window. We repeat this process a total of 100 times and average the error of the velocity field $\boldsymbol{u}$ and the fluctuations around the mean state $\boldsymbol{u}^{\prime}$.

The choice of the time step for comparison is based on the Lyapunov exponent found in wall-bounded turbulence \citep{keefe1992jfm,inubushi2015physreve,nikitin2018jfm}. We chose two time steps to compare: one at $t_1 =3 h/U_b$ equivalent to $t_1^{+} \approx 37$ and the second at $t_2 = 6 h/U_b$ equivalent to $t_2^{+} = 73$. The Lyapunov time $\tau_1$ is the inverse of the Lyapunov exponent, which in this case should lie between the value reported for the Couette minimal flow unit $\tau_{1}^{+} = 1/\lambda_1^{+} \approx 142,86$ \citep{inubushi2015physreve} and the one reported for the turbulent channel flow in the full size domain $\tau_1^{+} = 1/\lambda_1^{+} \approx 43,49$ \citep{nikitin2018jfm}. The first time $t_1^{+}$ is chosen below both Lyapunov times, but the second time $t_2^{+}$ is only below the value reported in \cite{inubushi2015physreve}. As our dimensions are in the minimal flow unit size, we believe that our model should behave similarly to the one presented in \cite{inubushi2015physreve}, as the Lyapunov exponent, scaled in wall units, is believed to be independent of the boundary conditions and Reynolds number \citep{nikitin2018jfm}.

\begin{figure}
  \centerline{\includegraphics[width=\textwidth]{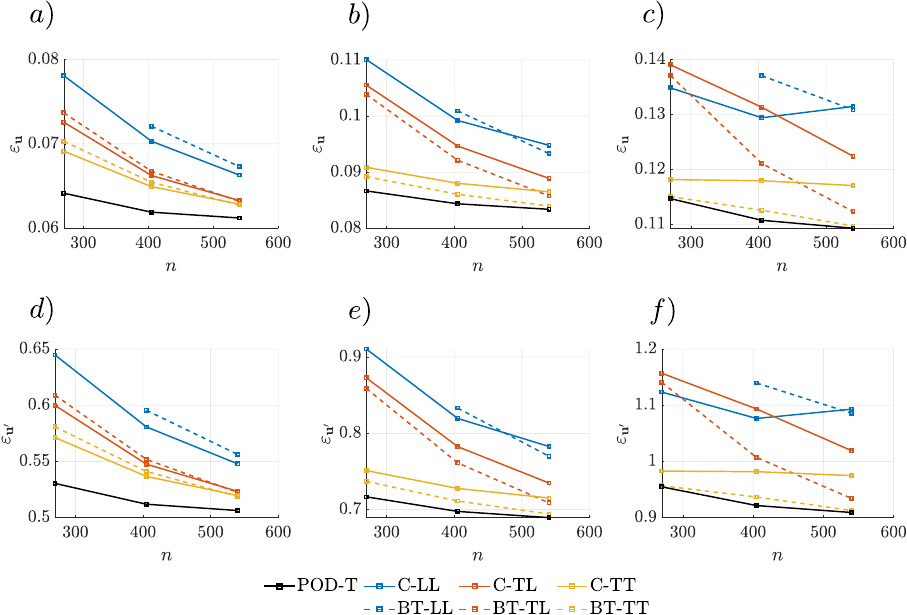}}% Images in 100% size
  \caption{Error for the full velocity $\boldsymbol{u}$ (a-c) and the fluctuations $\boldsymbol{u'}$ around the mean (d-f). In (a,d) $t_0^{+} = 0$, (b,e) $t_1^{+} \approx 37$ (c,f) $t_2^{+} \approx 73$. The errors are obtained after averaging 100 different random simulations of the ROMs. The models missing in the plots show blowing-up behavior in the calculation of statistics}
\label{fig:short_time_fore_full}
\end{figure}

The results are shown in Figure \ref{fig:short_time_fore_full}. The error in the velocity field in the initial condition confirms that POD is the best basis for projecting the field, as it uses data from the DNS. Bases using controllability modes (C-LL, C-TL, and C-TT) are better than balanced  truncation ones (BT-LL, BT-TL, and BT-TT) at the initial time, but we observe that as the number of modes increases, the differences can be negligible between approaches. Even though C-TL and C-TT use the same modal basis, they show different errors in the initial conditions, which can be explained using \ref{eq:projection_velocity}. The projection not only depends on the modal basis used, but also on the velocity fluctuations it represents. Mean-flow-based modal bases better project fluctuations onto the mean state than onto the laminar state, which explains the difference between the results.

Moving forward in time on the simulation, we observe an interesting phenomenon in Figure \ref{fig:short_time_fore_full}b,e where BT-LL, BT-TL, and BT-TT surpass their controllability modes counterparts (C-LL, C-TL, and C-TT). Balanced truncation modes perform better in the input-output dynamics, as demonstrated in several works (e.g., \citealt{rowley2005ijbc, ilak2008pof, barbagallo2009jfm}), given their accurate representation of transient growth, harmonic response, and structures that are dynamically relevant to the flow, regardless of their energy content. This trend is consistent if we simulate more time steps, as shown in Figure \ref{fig:short_time_fore_full}c,f with the model BT-TL surpassing C-TT when a sufficient number of modes are given. The comparison is studied until the error of the fluctuations $\varepsilon_{\boldsymbol{u}'} \approx 1$, which is achieved in Figure \ref{fig:short_time_fore_full}c,f.

The error for the full velocity field compared with the fluctuations is lower, as it can be compared in Figure \ref{fig:short_time_fore_full}a,b,c with d,e,f. The flow is predominantly influenced by the mean flow rather than by fluctuations. Then, a good representation of the mean flow is more important for the error calculation than an accurate representation of the fluctuations. The overall behavior if we change the comparison to the fluctuations results in the same conclusions as if we analyze the full velocity field.  

\begin{figure}
\centerline{\includegraphics[width=0.75\textwidth]{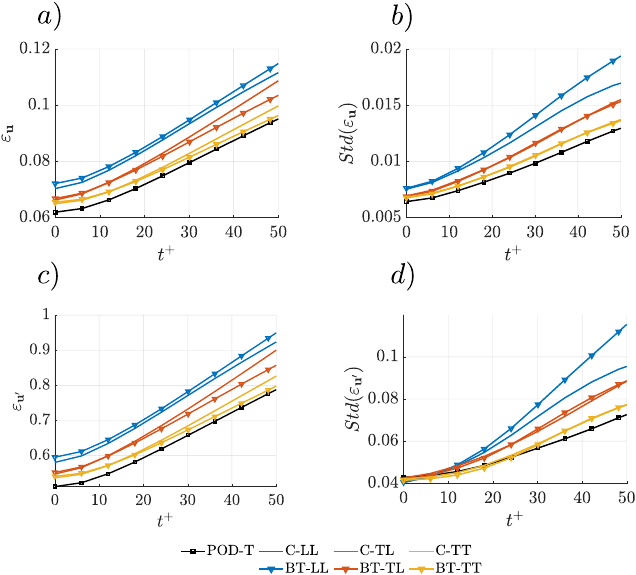}}% Images in 100% size
  \caption{Error for the fluctuations $\boldsymbol{u'}$ around the mean (a) and the full velocity $\boldsymbol{u}$ (c). Standard deviation error ($Std$) in (b) for $\boldsymbol{u'}$ and (d) for $\boldsymbol{u}$. The errors are obtained after averaging 100 different random simulations of the ROMs. Models use $n=405$ modes}
\label{fig:short_time_fore_time}
\end{figure}

An evolution of the error in both the velocity field and fluctuations around a mean state can be seen in Figure \ref{fig:short_time_fore_time} for models with $n=405$ modes. As appeared in Figure \ref{fig:short_time_fore_full}, ROMs from controllability modes and balanced modes start from similar initial error, and the only difference is due to the choice of modal bases and equations. These comparisons show that moving forward in time, BT-TL and BT-TT, in the average sense, have less error than their controllability counterparts, where this difference can be seen from $t^{+}\approx 30$ forward in time. The standard deviation error ($Std$) in Figure \ref{fig:short_time_fore_time} gives an insight of the variation of the error, where given the sensitivity to the initial condition, models start to have more deviation moving forward in time. The value of the error indicates that the average advantage of using BT-TT over C-TT is not necessarily valid when looking at single snapshots, where the situation of the error has to be assessed considering the initial condition. However, the improvement obtained using balanced truncation, albeit slight, is not related to a specific initial condition, displaying an average increase when several realisations are considered

% In the case of laminar-flow-based models with laminar-based modal bases (C-LL and BT-LL) the advantage of balanced modes over controllability is observed until $\varepsilon_{\boldsymbol{u}^{\prime}} \approx 1$, but for model that use mean-flow-based bases the balancing modes consistently achieve lower error independently of the equations used in the ROM. This trend is observed in both Figure \ref{fig:short_time_fore_time}a,b,  indicates that the mean-flow-based ROMs based on balancing modes produces lower error than other options like laminar-flow-based ROMs with laminar-flow modal bases or mean-flow modal bases.
\begin{figure}
\centerline{\includegraphics[width=.85\textwidth]{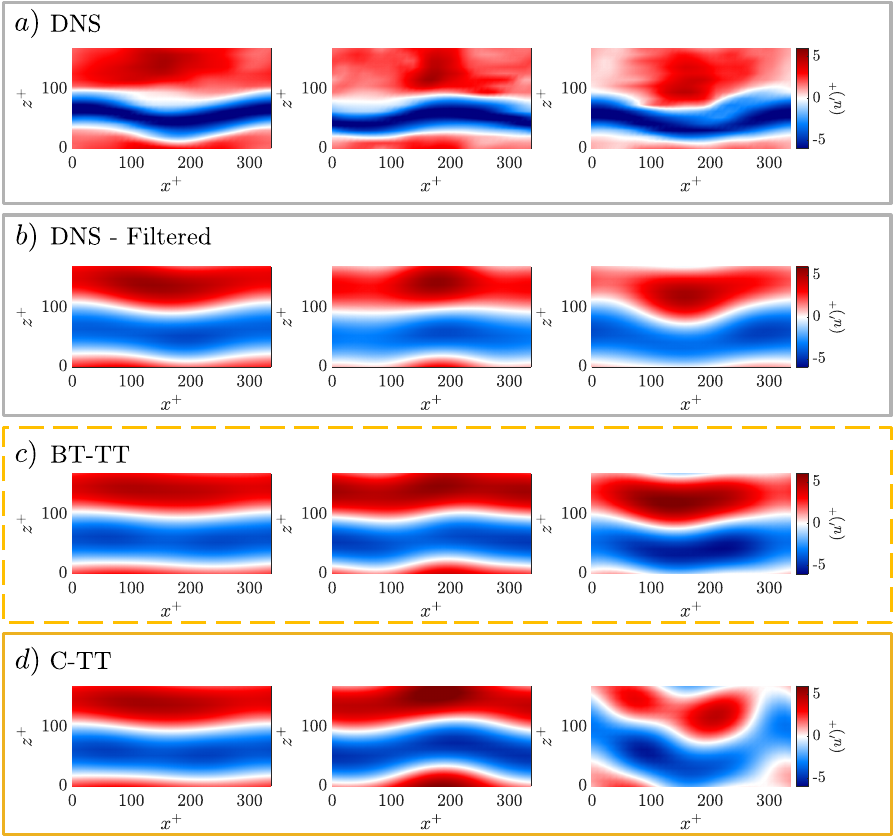}}% Images in 100% size
  \caption{Streamwise fluctuation velocity $u'$ in $y^+\approx20$ for three different times (left $t^{+} =0$, center $t^{+} \approx 42$, right $t^{+} \approx 85$) with (a) $\boldsymbol{u'}$ from DNS, (b) DNS filtered on the same wavenumbers of the ROM, (c) BT-TT, (d) C-TT. The ROMs use $n_y=36$ ($n=540$ modes)}
\label{fig:short_time_snaps_streamwise}
\end{figure}

The comparison of the streamwise velocity fluctuations is shown in Figure \ref{fig:short_time_snaps_streamwise}, where three different times from the simulation are plotted for an arbitrary initial condition in the location $y^{+} \approx 20$, focusing mainly on approaches based on equation-based ROMs around turbulent-state fluctuations (BT-TT, C-TT). The initial snapshots show a good agreement between the different models for a given initial condition. In time $t^{+} \approx 42$, the ROMs conserve a similar overall resemblance to the initial condition. Finally, at time $t^{+} \approx 85$, C-TT is significantly different from the DNS showing a streaky structure that is considerably more deformed than the BT-TT model.

\begin{figure}
\centerline{\includegraphics[width=.9\textwidth]{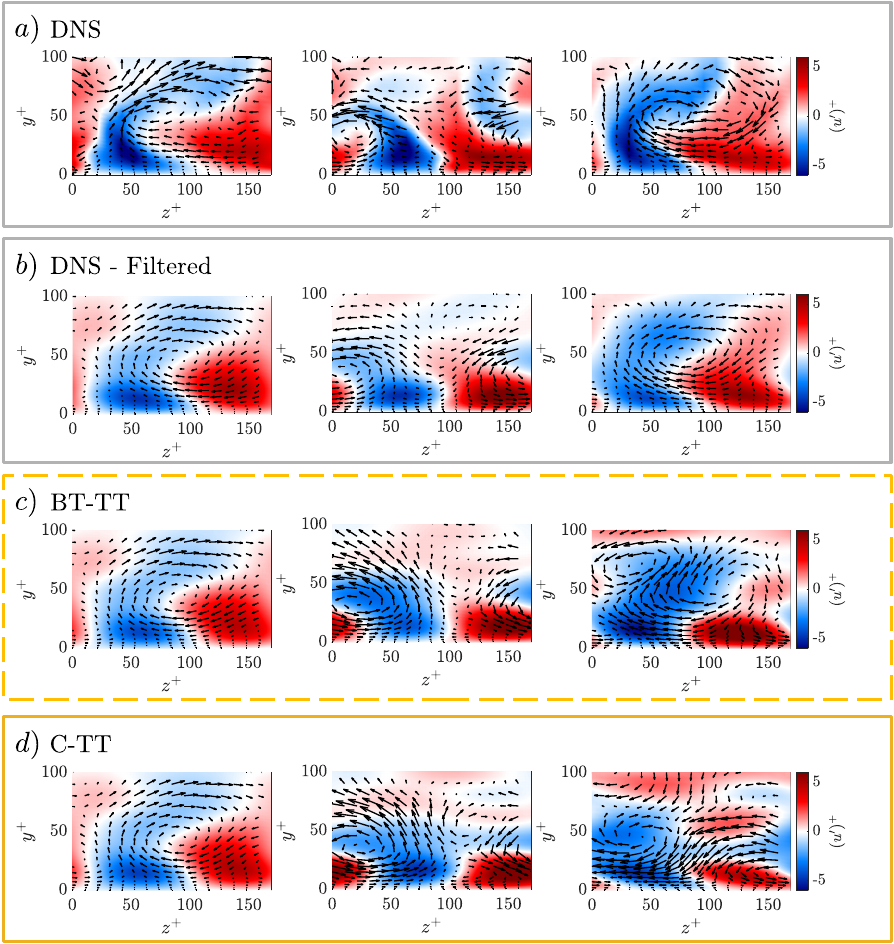}}% Images in 100% size
  \caption{Cross-section at $x^{+}\approx 168$  (left $t^{+} =0$, center $t^{+} \approx 42$, right $t^{+} \approx 85$) with (a) $\boldsymbol{u'}$ from DNS, (b) DNS filtered on the same wavenumbers of the ROM, (c) BT-TT, (d) C-TT. The ROMs use $n_y=36$ ($n=540$ modes)}
\label{fig:short_time_snaps_crosstream}
\end{figure}
\begin{table}
  \begin{center}
    \def~{\hphantom{0}}
  \begin{tabular}{lccccccc}
  & C-LL & BT-LL & C-TL & BT-TL & C-TT & BT-TT & POD  \\[8pt]
   Turbulent statistics & Weak & Weak & Fair & Fair & Strong & Strong  & Best \\
   Short-time forecasting & Weak & Weak & Weak & Strong & Weak & Strong & Best \\
   Data requirements & None & None & Mean flow & Mean flow & Mean flow & Mean flow & DNS snapshots
  \end{tabular}
  \caption{Comparison of the different ROMs. Details of the definitions in Table \ref{tab:roms_config}}
  \label{tab:rom_comparison}
  \end{center}
\end{table}
The cross-section of the channel is plotted in Figure \ref{fig:short_time_snaps_crosstream}. The cross-section is compared at the same times as Figure \ref{fig:short_time_snaps_streamwise} with an arbitrary initial condition from the DNS. The DNS snapshots indicate that the cross-stream flow appears in highly localized regions of values, whereas ROMs can reproduce it with reasonable accuracy in the initial snapshots. Looking further in $t^{+} \approx 42$, as in the previous Figure \ref{fig:short_time_snaps_streamwise}, both ROMs keep similar cross-stream velocity as streamwise velocity in the previous figure. In time $t^{+} \approx 85$, the cross-stream velocity where the BT-TT has a better representation of the vortical structures than C-TT.

\section{Conclusions}\label{sec:conclusions}
% key conclusion: change of equation and modal basis
% balancing truncation modes achieve greater dynamical accuracy with respect to controllability modes
% Extension to Poiseuille flow
% Applications, ECS, dissipation and understanding of SSP

In this work, we show how the equations of motion and modal bases influence the turbulent statistics convergence and short-time forecasting of ROMs based on Galerkin projection. The chosen case was a plane channel flow for its relevance to wall-bounded turbulence and the limited attention paid to equation-based ROMs for this specific configuration. In this mean-flow-based approach, the projection of Reynolds stresses is accounted for through a forcing term that can be obtained with the mean flow. A modification of the modal basis is presented to enforce the boundary conditions of plane channel flow. The change of both modal bases and equation of motion to the mean flow increases the accuracy of turbulence statistics and short-time forecasting in comparison with other ROMs based on laminar flow fluctuations \citep{cavalieri2022prf, zong2026jfm}.

Reduced-order modeling often relies on the ability to obtain data to build a modal basis that describes, in a reduced space, the dynamics of the turbulent flow. Equation-based modal bases allow building models without prior data and, as presented in this work, can perform similarly to POD-Galerkin models, only requiring knowledge of the mean state of the flow (C-TT and BT-TT). The study of different modal bases in a Couette flow was reported in \cite{zong2026jfm}, but the change in the equations from laminar-flow-based to mean-flow-based, presented in this work, can improve the performance of turbulent statistics beyond simply changing the modal basis of the ROM (e.g., C-TL compared with C-TT). An overall comparison is given in Table \ref{tab:rom_comparison}

ROMs for wall-bounded turbulence based on a modal basis derived from linearized equations were recently presented in several works (e.g., \citealt{cavalieri2022prf, mccormack2024jfm, zong2026jfm}), but the specific case of turbulent plane channel flow has not been previously studied. Here, we presented an extended framework that used a modified basis (Modified Stokes modes), in which the modal basis was chosen to have a null projection onto the pressure term of the equation of motion. This change allows the same mass flux to be sustained without setting a specific pressure gradient, as commonly POD-Galerkin based on the Navier-Stokes equation does in the literature \citep{omurtag1999tcfd, johansson2006compfluids}. However, unlike the cited works, the present Galerkin frameworks do not require numerical stabilisation.

The forecasting of equation-based ROM is presented for different modal bases and models for the dynamics. As with the turbulence statistics, mean-flow-based ROMs with mean-flow modal basis perform similarly to POD-Galerkin models, but here we see that for short-time forecasting, ROMs built from balanced truncation modes can achieve better short-time forecasting than their controllability modes counterparts in the case of mean-flow-based ROMs. This was achieved by averaging 100 different trajectories from the reference DNS and then simulating both the ROM and the DNS over a short time window.

This work generalizes the framework for ROMs based on linearized equations presented in \cite{cavalieri2022prf} for plane Couette flow to plane channel flow. This type of ROM can push the Reynolds number beyond that reported in previous work \citep{zong2026jfm} in a similar reduced space. The applications span from turbulence control \citep{maia2025jfm} to the computation of the invariant solutions representative of turbulence \citep{mccormack2024jfm}. The use of equation-based methods for building ROMs can yield results similar to those of data-driven methods like POD-Galerkin, where the use of mean-flow modal bases and equations is key to advancing into higher Reynolds numbers.

\begin{bmhead}[Acknowledgements]
We gratefully acknowledge E. Kracht for his helpful comments and insightful discussions.
\end{bmhead}

\begin{bmhead}[Funding]
Ian Addison-Smith acknowledges the financial support received from CAPES (Coordenação de Aperfeiçoamento de Pessoal de Nível Superior -
Brasil (CAPES) - Finance Code 001), under Move La America program, and ANID under scholarship ANID-Subdirección de Capital Humano/Magíster Nacional/2024-22241812. This work was also funded by ANID Fondecyt project 1250693.
\end{bmhead}

\begin{bmhead}[Declaration of interests]
The authors report no conflict of interest.
\end{bmhead}

\bibliographystyle{jfm}
\bibliography{jfm}

\end{document}